\documentclass[aps,prd,10pt,showpacs,amsmath,twocolumn, showkeys,onecolumn,floatfix,amssymb, preprintnumbers, nofootinbib, superscriptaddress]{revtex4}
\usepackage{graphicx}
\usepackage{comment}
\usepackage[usenames]{color}
\usepackage{bm}
\usepackage{ifpdf}
\usepackage{floatrow}
\usepackage{makecell}
\usepackage{subcaption}

\usepackage[normalem]{ulem}
\usepackage[dvipsnames]{xcolor}
\usepackage[utf8]{inputenc}
\usepackage{hyperref}
\hypersetup{
  pdfnewwindow=true,      
  colorlinks=true,        
  linkcolor=PineGreen,    
  citecolor=PineGreen,    
  filecolor=PineGreen,    
  urlcolor=PineGreen      
}
\begin{document}

\title{Polar magneto-optic Kerr and Faraday effects in finite periodic \texorpdfstring{$\mathcal{P}\mathcal{T}$}{PT}-symmetric systems}

\author{Antonio Perez-Garrido}
\email{Antonio.Perez@upct.es}
\affiliation{Departamento de F\'isica Aplicada,  Universidad Polit\'ecnica de Cartagena, E-30202 Murcia, Spain}

\author{Peng~Guo}
\email{pguo@csub.edu}

\affiliation{Department of Physics and Engineering,  California State University, Bakersfield, CA 93311, USA}

\author{Vladimir~Gasparian}
\email{vgasparyan@csub.edu}

\affiliation{Department of Physics and Engineering,  California State University, Bakersfield, CA 93311, USA}

\author{Esther~J\'odar}
\email{esther.jferrandez@upct.es}

\affiliation{Departamento de F\'isica Aplicada,  Universidad Polit\'ecnica de Cartagena, E-30202 Murcia, Spain}

\begin{abstract}

We discuss the anomalous behavior of the Faraday (transmission) and polar Kerr (reflection) rotation angles of the propagating light, in finite periodic parity-time ($\mathcal{P}\mathcal{T}$) symmetric structures, consisting of $N$ cells. The unit cell potential is two complex $\delta$-potentials placed on both boundaries of the ordinary dielectric slab.
It is shown that, for a given set of parameters describing the system, a phase transition-like anomalous behavior of Faraday and Kerr rotation angles in a parity-time symmetric systems can take place.
In the anomalous phase the value of one of the Faraday and Kerr rotation angles can become negative, and both angles suffer from spectral singularities and give a strong enhancement near the singularities. We also  shown that the real part of the complex angle of KR, $\theta^{R}_1$, is always equal to the $\theta^{T}_1$ of FR, no matter what phase the system is in due to the symmetry constraints. The imaginary part of KR angles  $\theta^{R^{r/l}}_2$ are  related to the $\theta^{T}_2$ of FR by parity-time symmetry.    Calculations based on the approach of the generalized nonperturbative characteristic determinant, which is valid for a layered system with randomly distributed delta potentials, show that the Faraday and Kerr rotation spectrum in such structures has several resonant peaks. Some of them coincide with transmission peaks, providing simultaneous large Faraday and Kerr rotations enhanced by an order one or two of magnitude. We provide a recipe for funding 
a one-to-one
relation in between KR and FR.

\end{abstract}

\maketitle

\section{Introduction}
The study of
the magneto-optic effects (Faraday rotation (FR) and Kerr rotation (KR)), has played an important role in the
development both of electromagnetic theory and atomic
physics. The 
magneto-optical materials that  exhibiting FR and KR are essential for optical communication
technology \cite{2005-1,2011-1,1982-1}, optical amplifiers \cite{2012-1,2000-1}, and photonic crystals \cite{prl-1,2009-N}.
In addition to this important application, the KR is also an extremely
accurate and versatile research tool and can be used to determine quantities
as varied as anisotropy constants, exchange-coupling strengths and
Curie temperatures (see, e.g., \cite{kr}.)

In
polar or magneto-optical Kerr effect, the magnetization of the system is in
the plane of incidence and perpendicular to the reflecting surface.
Reflection can produce several effects, including 1) rotation
the direction of light polarization, 2) introducing ellipticity into the reflected beam, and 3) changing
by the intensity of the reflected beam.

FR is similar to KR in terms of rotation and ellipticity and has 
a wide range of applications in various fields of modern physics, such as measuring magnetic field in astronomy \cite{longair_2011} and  construction of optical isolators for fiber-optic telecommunication systems \cite{9a}, as well as 
the design of optical circulators used in the development of microwave integrated circuits.
\cite{BERGER2003777,Turner:81,Firby:18}.

Note, that large Faraday and Kerr rotations are needed for all the applications mentioned. However, the standard method, based on increasing the sample size or applying a strong external magnetic field, is currently ineffective due to the small size of systems in which the de Broglie wavelength is compatible with the size of quantum devices. 
In other words, thin
film materials exhibiting a large FR angle should be desirable for
promote progress
optical integrated circuits.

A large enhancement of the FR and as well as a change in the sign of the FR can be obtained by incorporating several nanoparticles and their composites in nanomaterials, see e.g. Refs.~\cite{Uchida_2011,doi:10.1063/1.1625084,PhysRevA.89.023830}. A phase transition-like anomalous behavior of Faraday rotation angles in a simple parity-time $\mathcal{P}\mathcal{T}$-symmetric model of a regular dielectric slab was reported recently in Ref.\cite{GGJ-PLA}. In anomalous phase, the value of one of Faraday rotation angles turns negative, and both angles suffer spectral singularities and yield strong enhancement near singularities.

As for the enhancing of the KR, in which we are interested too, it is mainly related to
spin-orbit coupling strength \cite{kerr1}, to interference effects \cite{kerr2} and as well as by the plasma resonance of the free carriers of magnetic materials \cite{kerr3}.
As it was mentioned in Ref. \cite{2022}, 
with addition of a gold nano-disc to the periodic magnetic system, yields a strong wavelength-dependent enhancement of the KR.
Generally, the enhancement factor is expected to be less than three even for materials with high refractive index $\approx 2$, such as semiconductors with zero extinction coefficients in the near or mid infrared range (like tellurium
or aluminum gallium arsenide).

In this paper we aim to present a complete and quantitative theoretical description of the Faraday and Kerr complex rotations 
for an arbitrary one dimensional
finite periodic $\mathcal{P}\mathcal{T}$-symmetric system, consisting from (2N+1) cells
for some simple cases we give simple closed
form expressions, describing the FR and KR.

We illustrate that the Faraday and Kerr rotation angles of the  polarized light traveling through a $\mathcal{P}\mathcal{T}$-symmetric periodic structure display phase transition-like anomalous behaviors.

In one phase (normal phase), the FR and KR angles behave normally as in regular passive system with a positive permittivity, and stay positive all the time as expected.  In the second anomalous phase, the angle of FR and KR angles may change the sign and turn into negative.  
In addition, spectral singularities arise in the second anomalous phase, where the angles FR and KR increase strongly.
In this sense, $\mathcal{P}\mathcal{T}$-systems  seem to be a good candidate for constructing fast tunable and switchable polarization rotational ultrathin magneto-optical devices in a wide frequency range with a giant FR and KR rotations. Despite that the obtained results are,
in general, only suitable for numerical analysis. However, in
some simple cases approximate expressions can be derived
and a qualitative discussion is possible.

The paper is organized as follows. In Sec.~II the complex Faraday and Kerr effects are introduced and discussed for $\mathcal{P}\mathcal{T}$-symmetric unit cell with two complex $\delta$-potentials. We will assume that the strengths of two
Dirac $\delta$ functions $Z_1$ and $Z_2$ are arbitrary complex numbers.
The periodic system with $2N+1$ cells is is  discussed in Sec.~III. Followed by the discussions and summary in   Sec.~IV.

\section{General theory of Faraday and Kerr effects in \texorpdfstring{$\mathcal{P}\mathcal{T}$}{PT}-symmetric dielectric slabs}

\begin{figure}
\begin{center}
\includegraphics[width=0.9\textwidth]{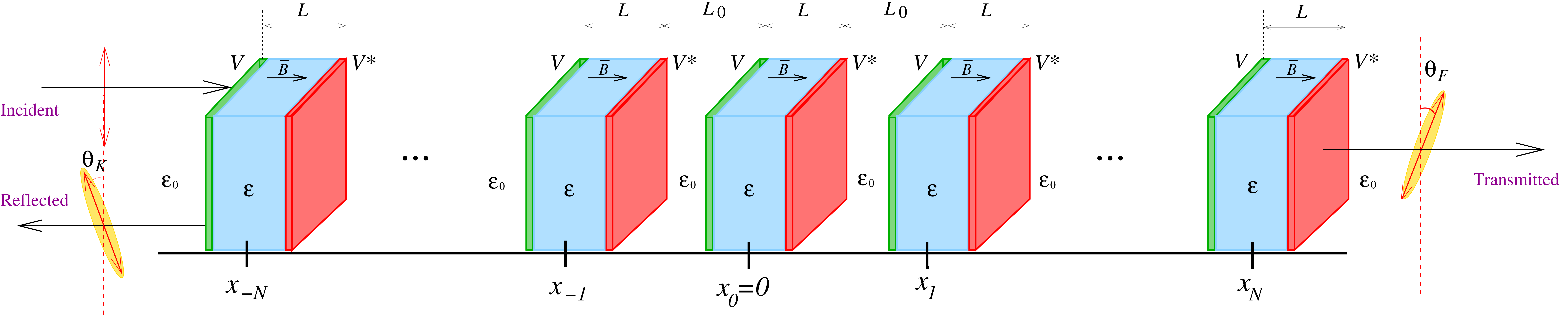}
\caption{Schematic of a one-dimesnionl $\mathcal{P}\mathcal{T}$-symmetric photonic heterostructure, consisting of $2N+1$ arbitrary number of slabs that are $\mathcal{P}\mathcal{T}$ -symmetric about $x_0 = 0$, that is $\epsilon (x)=\epsilon^{*} (-x)$. Each slab of the  photonic heterostructure, has two balanced complex tiny slabs placed at both ends of a real dielectric slab.
The green slab indicates the loss and the red slab indicates the gain region.
}\label{fig.sch}
\end{center}
\end{figure}

In this section, before discussing in detail the Faraday and Kerr effects in a simple unit cell—an ordinary dielectric slab with two complex $\delta$-potentials located at both boundaries (the unit cell located symmetrically about $x_0=0$ in Fig. 1), we present some details of the rotation angle calculation for an arbitrary one-dimensional dielectric permittivity profile $\epsilon (x)$. Later, we will impose the condition $\epsilon (x) = \epsilon^{*}(-x)$,
which guarantees the system is $\mathcal{P}\mathcal{T}$-symmetric, that its eigenstates are
real-valued solutions.
In such a $\mathcal{P}\mathcal{T}$-symmetric dielectric system with a finite spatial extension in $x$ direction (see Fig. 1), the permittivity of the system (as well as the single slab) has a balanced gain and loss.

Assume a linearly polarized electromagnetic plane wave with angular frequency $\omega$  enters the system from the left at normal incidence propagating along the $x$ direction. The polarization direction of electric field of incident wave is taken as the z-axis:  $\boldsymbol{ E}_0 (x)=   e^{i k_0 x} \hat{z}$,   where  $k_0=\frac{\omega}{c} \sqrt{\epsilon_0}$ stands for the wave vector and $\epsilon_0$ denotes the dielectric constant of vacuum.  A weak magnetic field $\boldsymbol{ B}$, which preserves the linearity of Maxwell's equations, is applied in the $x$-direction and is confined into the system, see Fig.1. The scattering of incident wave by the   system is described by Schr\"odinger-like equations, see e.g.  Refs.~\cite{Josh,PhysRevLett.75.2312},
\begin{equation}
\left [\frac{d^2}{d x^2}    + \frac{\omega^2 \epsilon_{\pm} (x)}{c^2}  \right ] E_\pm (x)  =0,
\end{equation}
where $E_\pm = E_y \pm i E_z$ are circularly polarized electric fields.   The $\epsilon_{\pm} (x) $ is defined,
\begin{equation}
\epsilon_{\pm} (x) = 
\begin{cases}  
\epsilon (x) \pm g, \ \ & \ \ x \in [-\frac{L}{2} - N(L+L_0),\frac{L}{2} + N(L+L_0)], \\
\epsilon_0, \ \ & \ \ \mbox{otherwise},
 \end{cases}
\end{equation}
where $(L,L_0,2N+1)$ stand for spatial extent of a unit cell, the spatial separation of neighbouring two cells and number of   cells, see Fig.~\ref{fig.sch}.  The   $g$ is the  gyrotropic vector along the magnetic-field direction. The external magnetic field $\boldsymbol{ B}$ is  included into the gyrotropic vector $g$   to make the calculations
valid for the cases of both external magnetic fields and magneto-optic materials.

When the reflection within the boundaries is important, the outgoing transmitted/reflected wave is generally elliptically polarized even without absorption, where the major
axis of the ellipse is rotated with respect to the original direction of polarization and the
maximum FR (KR) angle does not necessarily coincide with angular frequencies $\omega$ of light
at which zero ellipticity can be measured.
The real part of the rotation angle describes the change of polarization in linearly polarized light. The imaginary part describes the ellipticity of transmitted or reflected light.
Once we know the scattering matrix elements $r_\pm(\omega)$ and $t_\pm(\omega)$ of the one-dimensional light propagation problem,  e.g. the reflection and transmission amplitudes with an incoming propagating wave from left are defined by  
\begin{equation}
E_\pm (x) \rightarrow \begin{cases}
    \pm i \left [ e^{i k_0 x} + r_{\pm} (\omega) e^{- i k_0[ x+L +2N(L+L_0)]} \right ],       & \quad x \rightarrow - \infty, \\
    \pm i   t_{\pm} (\omega) e^{ i k_0   [x-L -2N(L+L_0)]  } , & \quad x \rightarrow + \infty.
  \end{cases}
\end{equation}
The two characteristic rotational parameters of transmitted light (magneto-optical measurements of complex Faraday angle) can be written as a complex form as (see, e.g., Refs. \cite{PhysRevLett.75.2312,Josh})
\begin{equation}
\theta^{T}_1  = \frac{\psi^{T}_+ - \psi^{T}_-}{2} , \ \ \ \  \theta^{T}_2 = \frac{1}{4} \ln \frac{T_+}{T_-}  ,\label {FR1}
\end{equation}
where $T_\pm$ and $\psi^{T}_\pm$ are the transmission coefficients and phase of transmission amplitudes,  $t_\pm=  \sqrt{ T_\pm} e^{i \psi^{T}_\pm} $,   of transmitted electric fields.  For weak magnetic field ($g\ll 1$), the perturbation expansion in terms of weak magnetic field can be applied. The leading order contribution can be obtained 
by expanding $\psi_{\pm}$ and $T
\pm$ around the refractive index of the
slab in the absence of the magnetic field B:
\begin{equation}
\theta^{T}_1    =
 \frac{g}{2n} \frac{\partial \psi^{T}}{\partial n} ,  \ \ \ \ 
\theta^{T}_{2}  =
 \frac{g}{4 n} \frac{\partial \ln{T}}{\partial n} ,\label{1mb}
\end{equation}
where
 $n= \sqrt{\epsilon}$ is the refractive index of the slab. The Kerr rotation complex angles are defined in a  similar way as in Eq.(\ref{FR1}). In the weak magnetic field, the leading order expressions can be written in the form
\begin{equation}
\theta^{R}_1    =
 \frac{g}{2n} \frac{\partial \psi^{R}}{\partial n} ,  \ \ \ \ 
\theta^{R}_{2}  =
 \frac{g}{4 n} \frac{\partial \ln{R}}{\partial n} ,\label{ref}
\end{equation}
where $R$ and $\psi^{R}$ are the reflection  coefficients and phase of reflection amplitudes in the absence of magnetic field B: $r(\omega) = \sqrt{R}e^{i \psi^R}$.

We remark that FR and KR angles are not all independent due to the constraints of  $\mathcal{P}\mathcal{T}$ symmetry. As mentioned in Ref.~\cite{PhysRevResearch.4.023083}, the parametrization of scattering matrix only requires three independent real functions in a $\mathcal{P}\mathcal{T}$-symmetric system: one inelasticity, $\eta \in [1, \infty]$, and two phaseshifts, $\delta_{1,2}$.  In terms of $\eta $ and  $\delta_{1,2}$,  the reflection and transmission amplitudes are given by
 \begin{equation}
 t  = t^r = t^l =\eta  \frac{ e^{2 i \delta_1 }+ e^{2 i \delta_2 }}{2}, \ \ \ \    r^{r/l} = \eta \frac{ e^{2 i \delta_1 }-   e^{2 i \delta_2 }}{2} \pm i \sqrt{\eta^2-1} e^{i (\delta_1+ \delta_2)},
 \end{equation}
where subscript $(r/l)$ are used to label amplitudes corresponding to two independent boundary conditions: right ($e^{i k_0 x}$) and left ($e^{- i k_0 x}$)  propagating incoming  waves respectively. Therefore we find relations:
\begin{equation}
 \sqrt{T}  =\eta \cos ( \delta_1 - \delta_2) , \ \ \ \  \psi^T = \delta_1 +  \delta_2 , \ \ \ \   \sqrt{R^{r/l}} =\left | \eta   \sin ( \delta_1 - \delta_2)   \pm  \sqrt{\eta^2-1} \right | ,  \ \ \ \ \psi^R   =  \psi^T + \frac{\pi}{2}, \label{inelastphaseeq}
\end{equation}
and the pseudounitary conservation
relations take place (see, e.g. Refs.\cite{Mostafazadeh_2014,Ahmed_2011,PhysRevA.85.023802}): 
\begin{equation}
|T-1|=\sqrt{R^{l} R^{r} }\label{ener} .
\end{equation}
The  FR and KR angles are given by
\begin{equation}
\theta^{T}_1 = \theta^{R}_1  =  \frac{g}{2n} \frac{\partial (  \delta_1 +  \delta_2 )}{\partial n},  \ \ \ \ 
\theta^{T}_{2}  =
 \frac{g}{2 n} \frac{\partial  }{\partial n}   \ln \left [ \eta \cos ( \delta_1 - \delta_2) \right ],  \ \ \ \  \theta^{R^{r/l}}_{2}  =
 \frac{g}{2 n} \frac{\partial  }{\partial n} \ln  \left | \eta   \sin ( \delta_1 - \delta_2)   \pm  \sqrt{\eta^2-1}   \right |.
\end{equation}
 The $\theta^{R^{r/l}}_{2} $ and $\theta^{T}_{2}$ are hence related by 
\begin{equation}
 \frac{\theta^{R^{r}}_{2} + \theta^{R^{l}}_{2}  }{2}  =     \frac{ T}{  T-1}  \theta^{T}_{2}  . \label{theta2RT}   
 \end{equation}
 We thus conclude that  only three FR and KR angles are independent due to the symmetry constraints.
The special case of zero inelasticity ($\eta=0$) thus represents the results for  real spatially symmetric dielectric  systems with   $\epsilon (x)=\epsilon (-x)$ and  $Im [\epsilon(x)  ]=0$,  hence 
\begin{equation}
\theta^{T}_{2}  \stackrel{Im [\epsilon(x)  ] \rightarrow 0 }{\rightarrow}
 \frac{g}{2 n} \frac{\partial  }{\partial n}   \ln    \cos ( \delta_1 - \delta_2)  ,  \ \ \ \  \theta^{R^{r/l}}_{2}  \stackrel{Im [\epsilon(x)  ] \rightarrow 0 }{\rightarrow}
 \frac{g}{2 n} \frac{\partial  }{\partial n} \ln \left |   \sin ( \delta_1 - \delta_2)   \right |  .
\end{equation}

\section{Unit cell: two complex \texorpdfstring{$\delta$}{delta}-potentials placed on both boundaries of the ordinary dielectric slab}

We first present some main results of FR and KR for a unit cell in this section, all the technical details can be found in Appendix \ref{detapproach}.   The properties of the spectral singularities are also discussed in current section, and we draw attention to the parameter ranges where a phase-like transition can take place for both Faraday and Kerr effects.   A simple $\mathcal{P}\mathcal{T}$-symmetric model for a unit cell is adopted in this work: two complex $\delta$-potential are placed at both ends of the dielectric slab,
\begin{equation}
\epsilon (x) = \epsilon + Z_1 \delta(x+ \frac{L}{2}) +Z_2 \delta (x - \frac{L}{2}), \ \ \ \ Z_1=V_1+i V_2, \ \ \ \ Z_2 =Z_1^*,
\end{equation}
where $L$ denotes the spatial extent of unit cell of dielectric slab and $\epsilon >0$  is  positive and real permittivity of slab.  The transmission $t_0(\omega)$ and reflection $r_0(\omega)$ amplitudes for the unit cell can be obtained rather straightforwardly by matching boundary condition method or using explicit form of characteristic determinant $D_2$ in Eq.(\ref{det2a}).

First of all, inserting Eq.(\ref{r21Ar21B}) in Eq.(\ref{det2}) and also using (\ref{t1a}) it is easy to see that $t_0(\omega)$,  phase $\psi^{T}$ and transmission coefficient $T_0$ for a unit cell are respectively given by
\begin{equation}
   t_0(\omega) = \sqrt{T_0} e^{i \psi^{T}} = \frac{   \csc \left(\frac{\omega n}{c} L\right) }{{\it \mathcal{R}}(\omega)- i {\it \mathcal{I}}(\omega)}, \ \ \ \ \psi^{T} = \tan^{-1}\left [ \frac {\mathcal{I} (\omega)}{\mathcal{R} (\omega)} \right ]   , \ \ \ \ T_0 = \frac{ \csc^2 \left(\frac{\omega n}{c} L\right)}{   {\it \mathcal{R}}^2(\omega) + {\it \mathcal{I}}^2 (\omega)},
   \label{t0}
\end{equation}
where
\begin{equation}
    {\it \mathcal{R}}(\omega) =  \cot \left(\frac{\omega n}{c} L\right) -   \frac{ \omega  V_1  }{c n}, \ \ \ \ {\it \mathcal{I}}(\omega)=  \frac{ \omega   V_1  }{c n_0}\cot{\left(\frac{\omega n}{c}L\right)}+ 	\frac{1}{2} ( \frac{n}{n_0}+\frac{n_0}{n})-  \frac{\omega^2}{2c^2 n_0 n}\left(V^2_1+V^2_2\right) . \label{ReIm} 
    \end{equation}
The $n = \sqrt{\epsilon}$ and $n_0 = \sqrt{\epsilon_0}$ denote the refractive index of the dielectric slab and vacuum respectively. We remark that unphysical units are adopted  throughout the rest of presentation: the length of slab $L$ is  used to sent up the physical scale,  $V_{1,2}$ and $\epsilon=n^2$ hence carry the dimensions of $1/L$ and $1/L^2$ respectively. The $\omega/c$ is   a dimensionless quantity.

Next the reflection amplitude $r^{r/l}_0(\omega)$ to the left/right of an individual cell can be  obtained conveniently  from the following relation related to the derivative of the transmission amplitude $t_0(Z_1, Z_2)$   with respect to $Z_1/Z_2$ located on the left/right border of the slab, see Ref. \cite{boris}:
\begin{equation}
r^{r}_0(\omega)=-
{i}\frac{cn_0}{\omega}{ \frac{\partial\ln t_0 (\omega)} { \partial Z_1}}-1, \ \ \ \ r^{l}_0(\omega)=-
{i}\frac{cn_0}{\omega}{ \frac{\partial\ln t_0 (\omega)} { \partial Z_2}}-1. \label{r}
\end{equation}

Hence we find
\begin{equation}
r^{r/l}_0(\omega) =\sqrt{R_0^{r/l}} e^{i \psi^R} =i\frac{   Q^{r/l} (\omega)}{\mathcal{R}(\omega)-i \mathcal{I}(\omega)}, \ \ \ \ \psi^{R} =  \tan^{-1}\left [ \frac {\mathcal{I} (\omega)}{\mathcal{R} (\omega)} \right ]  + \frac{\pi}{2}, \ \ \ \ R^{r/l} _0 = \frac{ \left[ Q^{r/l} (\omega) \right]^2}{{\it \mathcal{R}}^2(\omega) + {\it \mathcal{I}}^2 (\omega)}    ,  \label{r0}
\end{equation}
where 
\begin{equation}
Q^{r/l}(\omega)=\frac{\omega V_1}{cn_0}  \cot\left(\frac{\omega n}{c} L\right)+\frac{1}{2}
\bigg(\frac{n}{n_0}
-\frac{n_0}{n}\bigg)
\pm\frac{\omega V_2}{cn}  -\frac{\omega^2}{2c^2n_0n}\left( V^2_1+V^2_2\right)\label{Q}.
\end{equation}
Note, that in case of $n_0$ = $n$ we recover the result of a reflection amplitude from a simple diatomic   system, discussed in Ref. \cite{peng22}.
  It is easy to verify that the phase of the reflection amplitude indeed coincides with the phase of the transmission amplitude  as previously discussed. Later, in the next subsections we used these expressions to illustrate a number of quite general features of Faraday and Kerr rotations in $\mathcal{P}\mathcal{T}$-symmetric periodic systems.

A simple inspection of the Eq.(\ref{t0}), show that replacing $\omega$ with $-\omega$ does not affect $t_0(\omega)$, which means that the transmission is equal for the left-to-right and right-to-left
scattering, that is $t_0^{l}(\omega) = t^{r}_0({-\omega})\equiv t_0(\omega)$. The situation is somewhat more complicated in the case of the reflection amplitude in Eq.(\ref{r0}).
Simultaneous sign change of both $\omega$ and $V_2$ is required to satisfy the condition $r_0^{l}(-\omega, - V_2)= r^{r}_0(\omega, V_2)$. There   in fact are indeed the  general properties of $\mathcal{P}\mathcal{T}$ systems, see e.g. Eq.(B33) in Ref.~\cite{PhysRevResearch.4.023083}.

  \begin{figure*}
 \centering
 \begin{subfigure}[b]{0.49\textwidth}
\includegraphics[width=0.99\textwidth]{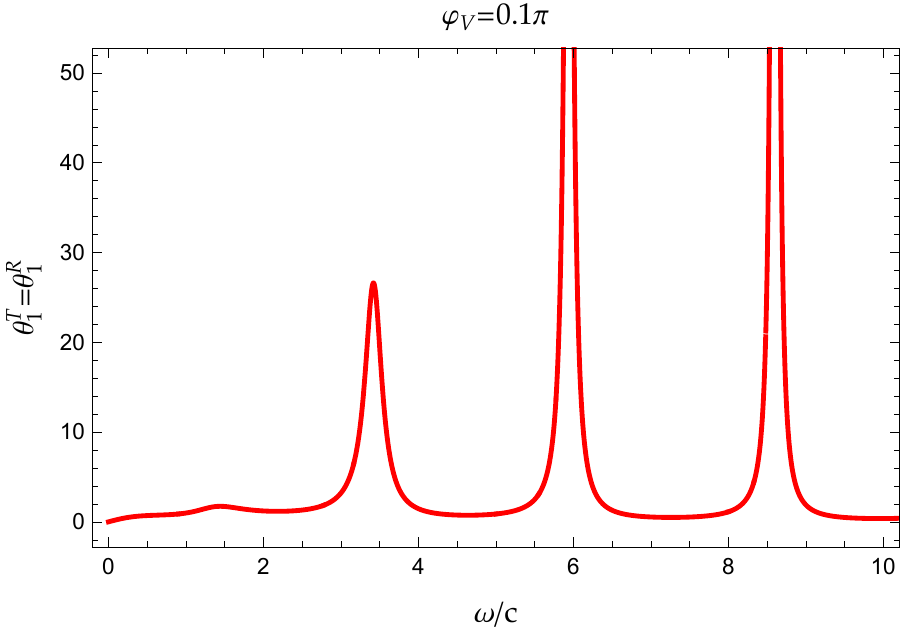}
\caption{  }\label{theta1plot1}
\end{subfigure} 
\begin{subfigure}[b]{0.49\textwidth}
\includegraphics[width=0.99\textwidth]{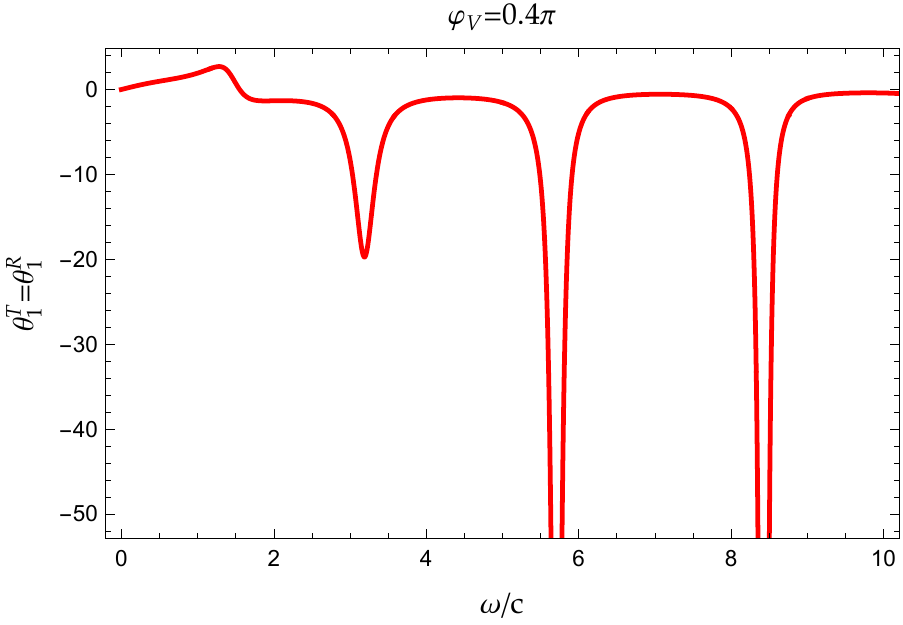}
\caption{    }\label{theta1plot2}
\end{subfigure}
\begin{subfigure}[b]{0.49\textwidth}
\includegraphics[width=0.99\textwidth]{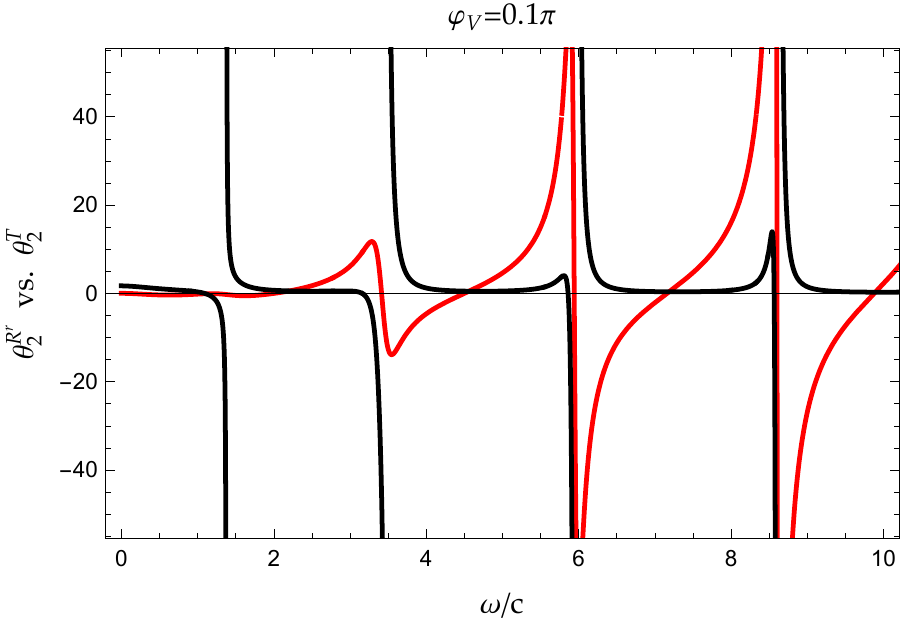}
\caption{ }\label{theta2plot1}
\end{subfigure}
\begin{subfigure}[b]{0.49\textwidth}
\includegraphics[width=0.99\textwidth]{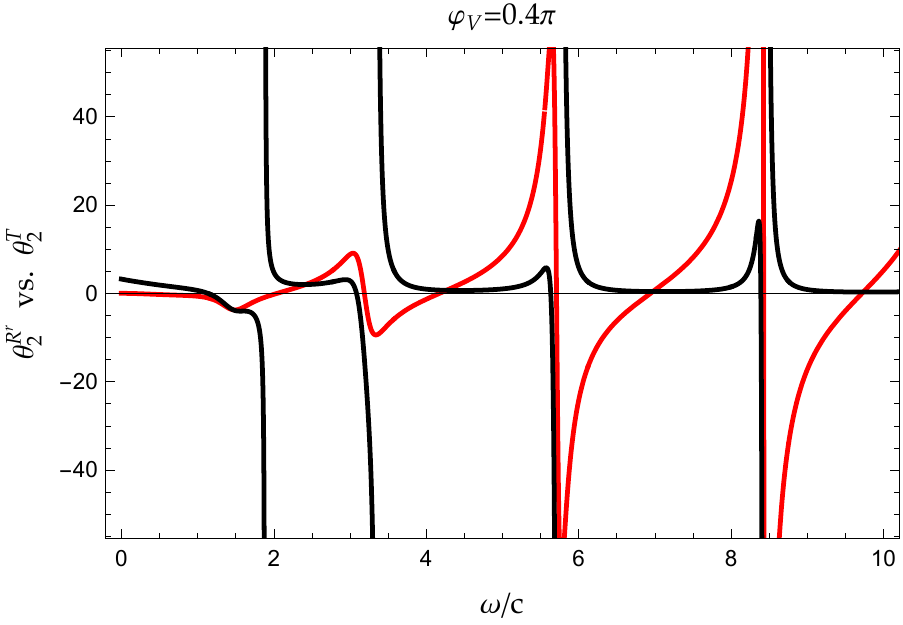}
\caption{  }\label{theta2plot2}
\end{subfigure}
 \caption{Faraday   and Kerr   rotation angles  $\theta^{T}_{1,2}$ and $\theta^{R^r}_{1,2}$   versus $\frac{\omega}{c}$ are shown for  two values of $\varphi_{V} = 0.1 \pi $ and $0.4\pi$ for $\mathcal{P}\mathcal{T}$-symmetric unite cell: (a)  $ \theta^T_1= \theta^R_1$ for $\varphi_V = 0.1\pi$; (b) $ \theta^T_1= \theta^R_1$ for $\varphi_V = 0.4\pi$; (c) $\theta^{R^r}_2 $ (solid red) vs. $\theta^{T}_2$ (solid black) for $\varphi_V = 0.1\pi$; (d) $\theta^{R^r}_2 $ (solid red) vs. $\theta^{T}_2$ (solid black) for $\varphi_V = 0.4\pi$.  The rest of parameters are taken as:       $|V|=1$,   $L=0.8$, $n=\sqrt{2}$ and $n_0=1$.  \label{fig.2}}
 \end{figure*}

\subsection{Spectral singularities in a unit cell}\label{specsingunitcell}

We now turn to a closer investigation of the spectral singularities for FR and KR angles.  Spectral singularities are spectral points belonging to non-Hermitian Hamiltonian operators with $\mathcal{P}\mathcal{T}$-symmetry, characterized by real energies. At these energies, the reflection and transmission coefficients tend to infinity, i.e., they correspond to resonances having zero width. 
Interesting to note that a slight imbalance between gain and loss regions,  can change the shape of the transition from zero width to the symmetric shape of the "bell curve" (for more details see Ref.~\cite{GGJ-PLA}).

For our model and for FR and KR rotational effects, spectral singularities arise when both
conditions, $\mathcal{R}(\omega) = 0$ and $\mathcal{I}(\omega) = 0$, are satisfied  simultaneously, see Eq.(\ref{ReIm}).
 By solving Eq.(\ref{ReIm}) for $\cot (k L)$
and $\omega$ one obtains straightforwardly
\begin{equation}
    \bigg(\frac{  \omega_{cr}  |V|  }{c }\bigg)^2\cos{2\varphi_{V}}+n^2+n^2_0=0 , \ \ \ \ |V|=\sqrt{V_1^2+V_2^2},
\end{equation}
where $\tan(\varphi_{V})=\frac{\mathcal{I} (\omega_{cr} )}{\mathcal{R} (\omega_{cr} )}$.
The condition necessary for the existence of a solution 
of the spectral singularities exist only when  the transmission phase is in the range 
$\varphi_{V} \in [\pi/4,\pi/2]$. Hence the critical value of $\omega_{cr}$ is defined  as
\begin{equation}
\omega_{cr}=\frac{c}{|V|}\frac{\sqrt{n^2+n^2_0}}{\sqrt{|\cos{2\varphi_{V}}|}}, \label{cr}
\end{equation}
provided that
\begin{equation}
\cot \bigg(\frac{\omega_{cr}}{c} nL\bigg) = \frac{\sqrt{n^2+n_0^2}}{n}
\frac{\cos{\varphi_{V}}}{\sqrt{|\cos{2\varphi_{V}}|}}. \label{cr1} \end{equation}
 For a fixed $|V|$, as follows from Eq.(\ref{cr}) the solutions of spectral singularities can only be found in a finite range:  $\varphi_{V} \in [\frac{\pi}{4}, \varphi_c]$, where $\varphi_c$ stands for upper bound of range. Hence as $\varphi_{V}$ approaches lower bound of range at $  \frac{\pi}{4}$, the spectral singularity solution occurs at large frequency: $\omega \rightarrow \infty$. When $\varphi_V$ is increased, the solution of  spectral singularity   moves toward lower frequencies.  As $\varphi_{V}   $ approaches the upper bound of range at $ \varphi_c$,   the spectral singularity solution thus reaches its lowest value. The graphical illustration of the distribution of spectral singularities can be found in Fig.2 in Ref.~\cite{GGJ-PLA}.

\subsection{Faraday and Kerr rotation: transmitted and reflected light}

 A phase transition-like anomalous behavior and properties of Faraday rotation angles in a simple   $\mathcal{P}\mathcal{T}$-symmetric model with two complex $\delta$-potential placed at both boundaries of a regular dielectric slab was most recently reported in Ref.\cite{GGJ-PLA}.
 Let us recall the essential features of the FR and then focus our attention on the KR effect. In a  $\mathcal{P}\mathcal{T}$-symmetric systems a phase transition-like anomalous behavior of Faraday rotation angle take place. In this phase, one of Faraday rotation angles turns negative, and both angles yield strong enhancement near spectral singularities.

 As the consequence of $\mathcal{P}\mathcal{T}$ symmetry constraint, the phase of reflected amplitude $\psi^{R}$ from left coincides with the phase of the transmission amplitude $\psi^{T}$, see Eq.(\ref{inelastphaseeq}). Hence the real part of the complex angle of KR, $\theta^{R}_1$, is always equal to the $\theta^{T}_1$ of FR, no matter what phase the system is in.   In this sense, the
situation is similar to the 
passive symmetric system, where is always 
$\theta^{R}_1=\theta^{T}_1$. 
It is interesting to note that Eq.(\ref{ReIm}) is invariant under the symmetry transform: $V_2 \rightarrow -V_2$. This is a manifestation of the fact that the phase of reflected amplitude $\psi^{R}$ and as well as the Kerr rotation angle for the right incident light preserve the same behaviour, although the strengths of the right and left $\delta$-potentials on the boundaries are not equal to each other (more precisely, they are complex conjugate to each other). 
 The mentioned asymmetry should lead to different left-to-right  and right-to-left reflection amplitudes (see, e.g., \cite{opensyst1}) and does not affect physical quantities $\theta^{R}_1$ and $\theta^{T}_1$, which are related to the phase accumulated during the process of  reflection and transmission and as well as to the density of states. 
However, this asymmetry will affect $\theta^{R}_2$ and $\theta^{T}_2$, and they will no longer be equal to each other, see Fig.~\ref{fig.2}(c) and Fig.~\ref{fig.2}(d). This is consistent with the general statement that the Faraday and Kerr rotation profiles are very different from the corresponding curves describing ellipticities. In addition, symmetry constraint also yields the wavelength dependence of Faraday and Kerr  ellipticity  $\theta^{T}_2$ and $\theta^{R^{r/l}}_2$ shown in  Eq.(\ref{theta2RT}).

Here we would like to add a few more brief comments to emphasize that upon closer look at Fig.\ref{fig.2} reveals some details of the 
similarities between curves that are relevant to our further discussion.

Firstly, the Faraday (Kerr) rotation local maximum/minimum (see Fig.~\ref{fig.2}) coincide with the local peak on the ellipticity curves with some accuracy. The ellipticity, at that frequencies, approaches zero non-linearly, becomes zero (linearly polarized light), and then the resulting polarization reverses its original direction.

Secondly, ellipticity (imaginary part
the spectra) for $\theta^{T}_2$ and $\theta^{R}_2$ depend little on frequency and are close to zero in almost the entire frequency range, except for some regions associated with the maximum/minimum or spectral singularities of the Faraday rotation and Kerr rotation.

The questions discussed above can be straightforwardly generalized for the periodic $\mathcal{P}\mathcal{T}$-symmetric system. This will be done in the next section. We will show that   the anomalous effect, similar to a phase transition, occurred more often due to the complex structure of the transmission and reflection amplitudes.

\subsection{ Limiting cases }

 The phase transition-like behavior of $\theta^{T}_1$  for two limiting cases $ (|V| \rightarrow \infty$ and $V_1 \rightarrow  0$) was discussed in Ref.~\cite{GGJ-PLA}. It was shown that in the case of  $|V| \rightarrow \infty$ the sign of $\theta^{T}_1$ is completely determined by $\varphi_{V}$. As for the case $V_1=0$ ($\varphi_{V} \rightarrow  \frac{\pi}{2}$), then again for the given parameters of the problem
the anomalous negative behavior of $\theta^{T}_1$ is illustrated analytically.
The latter case, that is a $\mathcal{P}\mathcal{T}$-symmetric optical lattice with a purely imaginary scattering potential has been discussed in detail in a number of investigations both theoretically and experimentally, see, e.g.
Refs.~\cite{opensyst1} and references therein.

\subsubsection{ \texorpdfstring{$|V| \rightarrow \infty$}{|V|->infinity} }
The situation is slightly different for Kerr rotation. In the same limiting case
$|V| \to \infty$, given that $ \frac {n\omega L}{c}\ne \pi l$, we can show that $\theta^{R}_{2} \propto  \frac{1}{ |V |^2} $, hence the ellipticity is almost zero for all frequencies  excluding $\frac {n\omega L}{c} =\pi l$ where $l \in \mathbb{Z}$ and the reflected light remains linearly polarized.   At the discrete values of $\omega/c = \frac{\pi l}{n L}$ that yield the location of the resonance poles,  $\theta^{R}_{2} $   display sharp peak with narrow resonance width. It reflects the fact that the reflected light is again linearly polarized but rotated 90 degrees from the initial direction.

\subsubsection{ \texorpdfstring{$ V_1 \rightarrow 0$}{V1->0} }
\begin{figure}
\includegraphics[width=0.49\textwidth]{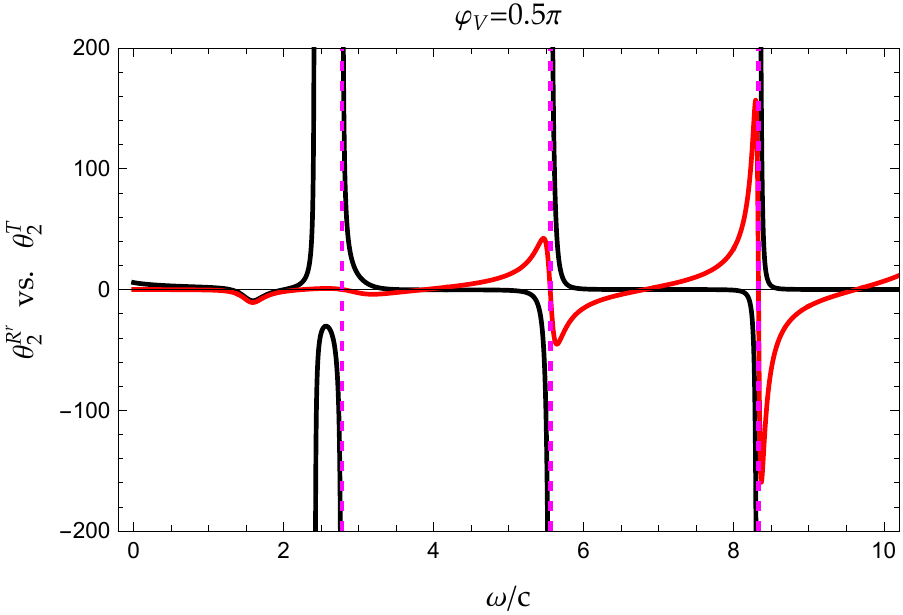}
 \caption{Plot of Faraday $\theta^{T}_2$ (solid red) and Kerr $\theta^{R^r}_2$ (solid black) rotation angles  are shown for  $|V|=1$,  $\varphi_{V}=\frac{\pi}{2}$, $L=0.8$, $n=\sqrt{2}$ and $n_0=1$ for $\mathcal{P}\mathcal{T}$-symmetric unite cell, respectively. The resonant frequencies at $\frac{\omega_l n}{c}L=\pi l, \   l=1,2, \cdots$ are plotted as dashed pink vertical lines.   } \label{fig.3}
 \end{figure}

Bound state solutions of the Schr{\"o}dinger equation for a $\mathcal{P}\mathcal{T}$-symmetric
potential with Dirac delta functions was study in Ref.\cite{uncu}. 
In Ref.\cite{GGJ-PLA} it was pointed out that despite the fact that the expression for $\theta^{T}_1$ is valid for the case $V_1 \rightarrow  0$, it can still explain not only the sign change of $\theta^{T}_1$ ($\theta^{R}_1$) in Fig.\ref{fig.2} (a) where $V_1 \ne 0$, but also explain existence
first local maximum. It is clear that further features of the $\theta^{T}_{1}$ ($\theta^{R}_1$) in Fig.~\ref{fig.3}(a) near the frequencies of spectral singularities, is related to the behavior of $T(\omega)$.

As for the imaginary portion of Kerr effect $\theta^{R^{r/l}}_{2}$, it is straightforward to show that in the same limit of $V_1 \rightarrow  0$ the $\theta^{R^{r/l}}_{2}$ reads  
 \begin{equation}
 \theta^{R^{r/l}}_{2}\stackrel{V_1 \rightarrow 0}{\rightarrow}\frac{g}{2 nQ  (\omega)}\left [  \frac{1}{2}\bigg(\frac{1}{n_0}+\frac{n_0}{n^2}\bigg)\mp \frac{\omega}{cn^2} V_2\bigg(1 \mp \frac{\omega}{2cn_0} V_2\bigg)
 +\frac{R^{r/l}_0}{Q^{r/l} (\omega) }\bigg(\frac{\omega L}{c}\frac {\cot (\frac{n\omega}{c} L)}{\sin^2(\frac{n\omega}{c} L)}-\frac{n^2-(n^2_0-\frac{\omega^2}{c^2} V^2_2)^2}{4n^2_0n^3}\bigg)\right ]
,\label{1mba}
\end{equation}
where the reflection coefficient $R^{r/l} _0$ is given by Eq.(\ref{t0}) and $Q^{r/l} (\omega)$ is defined by Eq.(\ref{Q}).    The dependence of the imaginary part of the Kerr rotation $\theta^{R^{r}}_2$ (solid black line) on $\frac{\omega}{c}$ for $V_1 \rightarrow  0$ is illustrated in Fig.~\ref{fig.3}. A number of basic features of $\theta^{R}_2$ can be observed even in this simplest case of $V_1 \rightarrow  0$. One of the key features is the single resonant peak that show up clearly when $R_0\rightarrow \infty$, see Eq.(\ref{1mba}). As mentioned above the resonance frequencies are spectral singularities when both
conditions, ${\it \mathcal{R}}(\omega)=0$ and ${\it \mathcal{I}}(\omega)=0$, are satisfied simultaneously. 
In the particular case of $V_1 \rightarrow  0$ there is only one
 $\omega_{cr}$ that can be directly calculated from Eq.(\ref {cr}) by putting $\varphi_{V}=\frac{\pi}{2}$:
$\omega_{cr}=\frac{c\sqrt{n^2+n^2_0}}{V_2}$. 
The second condition $\cot\bigg(\frac{\omega_{cr}}{c}nL\bigg)=0$
can be satisfied 
by choosing
the appropriate value of length is $L=0.8$ (the system parameters are: $n_0=c=V_2 =1$, $n=\sqrt{2}$ and $\omega_{cr}=\sqrt 3$). Other maxima or minima in the Kerr rotation, located near the resonant frequencies, are associated with multiple reflections from the boundaries and are located at $\frac{\omega_l n}{c}L=\pi l, \   l=1,2, \cdots$ (see vertical pink lines in Fig.\ref{fig.3}).

Repeating similar calculations leading to Eq.(\ref{1mba}), 
we arrive at an explicit expression for ellipticity $\theta^{T}_{2}$ 
for Faraday rotation for this simplest case with a purely imaginary potential:
\begin{equation}
 \theta^{T}_{2}\stackrel{ V_1 \rightarrow 0}{\rightarrow}
\frac{g}{2 n}{\cot(kL)}\frac{\omega}{c}L\left [ T_0\left (1-
\frac{{\sin^2(kL)}}{\cot(kL)}\frac{c}{\omega L}\frac{n^2-(n^2_0-\frac{\omega^2}{c^2} V^2_2)^2}{4n^2_0n^3}\right )-1
\right ].\label{1mby}
\end{equation}
We observe that the smoothed maxima and minima 
that appeared around the zeros of $\sin{kL}$
at $\frac{\omega_l n}{c}L=\pi l, \  l=1,2, \cdots$   coincides  with maxima and minima of $\theta^{R}_{2}$
and  associated with multiple reflections from the boundaries,  see  e.g. vertical yellow lines in Fig.\ref{fig.3}. 
Secondly, the large value of $\theta^{T}_{2}$ at $3 \pi /2$ is related to the frequency of the spectral singularity $\omega_{cr}=\frac{c\sqrt{n^2+ n^ 2_0}}{V_2}$, where $T_{0} \rightarrow \infty$.

The physical background of the relatively simple mathematical structure of the Faraday rotation angle  $\theta^{T}_{1}$
 on the frequency of comparison Kerr ellipticity ($\theta^{R^{r/l}}_{2}$ is that in the first case the rotation maximum is direct proportional to the optical anisotropy (for example, the larger ( $n_{+} - n_{-}$ ), the larger is $\theta^{T}_{1}$. However,
 the maximisation of $\theta^{R^{r/l}}_{2}$
is, not so straightforward, since anisotropy indices
are mixed (see, e.g, Ref. \cite{nemec} and  and references therein).

\section{Periodic system with \texorpdfstring{$2N+1$}{2N+1} cells}

It is known that when  the wave
propagation through a medium is described by a differential equation of second order, the expression
for the total transmission from the finite periodic system for any waves (sound and electromagnetic) depends on the unit cell transmission,
the Bloch phase and the total number of cells. 
As an example of collective interference effect, let us mention the intensity distribution from $N$ slits (diffraction due to $N$-slits), as well as the formula that describes the  Landauer's 
resistance of a one-dimensional chain of periodically spaced $N$ random scatterers.
In both cases, the similarity of the results is obvious. However, the physics behind these results is completely
different both in spirit and in details.
In analogy to Hermitian Hamiltonian, one can expect interference effect holds also for a non-Hermitian Hamiltonian system.
In this sense it is a natural result for a $\mathcal{P}\mathcal{T}$-symmetric system that 
a somewhat similar formula for transmission and reflection amplitudes appears, for example, in Refs.~\cite{periodic1,periodic2,li,peng22}.
The infinite periodic $\mathcal{P}\mathcal{T}$-symmetric structures, 
because of unusual properties, including 
the band structure, Bloch oscillations,
unidirectional
propagation and enhanced sensitivity, 
are of special interest and are presently the subject of intensive ongoing research (see e.g., Refs.~\cite{ben,shin,Mus,mid1} and references therein).
However, the case of scattering
in a finite periodic systems composed of an arbitrary number of
cells/scatters has been less investigated, despite 
that any open quantum systems generally consist of a finite system coupled with an infinite environment.

In many studies, 
to describe quantitatively, both amplification and absorption in periodic
$\mathcal{P}\mathcal{T}$-symmetric systems, the transfer matrix method is used. The latter, can be reduced to the evolution of
the product of transfer matrices
of complex, but identical unit cells, and using
the classical Chebyshev identity 
get the final result.

In the following, we present the amplitudes of transmission and   reflection form the left and right sides of the incident wave based on the characteristic determinant approach, the technical details are given in Appendix \ref{detapproach}. The latter, in principle, is compatible with the transfer matrix method and is convenient for both numerical and analytical calculations.

\subsection{Amplitudes of transmission and reflection form left and right}

We now turn to a closer investigation of the Faraday and Kerr rotations for various parameter ranges of our $\mathcal{P}\mathcal{T}$-periodic symmetric system that
consists of $2N+1$ cells, see Fig.\ref{fig.sch}. 
Following Refs.~\cite{Esther-1997,peng22} and also see Appendix \ref{detapproach},   a generic expressions for the transmission and left/right reflection amplitudes for the $\mathcal{P}\mathcal{T}$ can be presented as: 
\begin{align}
t (\omega)& =\frac{e^{-ik_0L_0}}{\cos(\beta(2N+1)\Lambda)+ i Im \left [\frac{e^{- i k_0L_0 } }{ t_0 (\omega)}\right] 
\frac{\sin(\beta(2N+1) \Lambda)}{\sin (\beta \Lambda)}}, \label{tandrexpress}
\end{align} 
where $k_0=n_0\frac{\omega }{c}$ and $k=n\frac{\omega}{c}$ are the wave
vectors in the respective medium.
The quasi-momentum $\beta$ is the Bloch wave vector 
of the infinite periodic system with unit cell length or spatial periodicity  $\Lambda=L_0+L$: 
\begin{equation}
\cos ({\beta \Lambda})\equiv  {\it Re} \left [\frac{ e^{-i k_0 L_0  } }{t_0 (\omega)} \right ] = \sin ( k L)  \left [ \cos (k_0 L_0)    \mathcal{R}(\omega) -  \sin (k_0 L_0)     \mathcal{I}(\omega)  \right] .
\end{equation}  please confirm two equations in blue. 
The left/right reflection amplitude can be written in the form \cite{peng22,Esther-1997}
 \begin{align}
 \frac{ r^{(r/l)} (\omega) }{t(\omega) } & = \left [  \frac{ r^{(r/l)}_0 (\omega) }{t_0(\omega)}  \right ] \frac{\sin (\beta (2N+1) \Lambda)}{\sin (\beta \Lambda) }     , \label{tandrexpress1}
\end{align}
where $t_0 (\omega)$  and ${ r^{(r/l)}_0 (\omega) }$  are the 
transmission and reflection amplitudes for a single cell ($N=0$) that are given in   Eq.(\ref{t0}) and Eq.(\ref{r0}) respectively.

An important feature of expressions (\ref{tandrexpress})
and (\ref{tandrexpress1}) is that both contain factor $\frac{\sin (\beta (2N+1) \Lambda)}{\sin (\beta \Lambda)}$ which naturally occur in Hermitian one-dimensional finite periodic systems due to interference or diffraction effects and reflects
a combine effect of all $2N+1$ cell. The appearance of this factor in non-Hermitian systems is 
highly non-trivial from the view 
of the usual probability conservation property 
for Hermitian systems (the reflection and transmission coefficients must sum to unit   in either classical or quantum
mechanical regimes)
or unitary scattering matrix theory. However, in Refs. \cite{periodic1,peng22} a simple closed form expressions is obtained for the total transmission and reflection (left/right) amplitudes from a
lattice of $N$ cells. As pointed out in Refs.~\cite{peng22}, the   transmission and reflection amplitudes for a periodic many  scatters system are   related to single cell   amplitudes in a compact fashion. 
This is
intimately connected with the fact that the factorization of short-range dynamics in a single cell and long-range collective effect of   periodic structure of entire system:   the short-range interaction dynamics that is described by single cell scattering amplitudes and  the $\beta$  represents the collective mode of entire lattice. The occurrence of factorization  of  short-range dynamics and long-range collective mode   has been known   in both condensed matter physics and nuclear/hadron physics.  In the cases such as particles interacting with short-range potential in a periodic box or trap,  where two physical scales,  (1) the short-range particles dynamics and (2) long-range geometric effect due to the  periodic box or trap,  are clearly separated.   The quantization conditions are given by a compact formula   that is known as Korringa–Kohn–Rostoker  (KKR) method  \cite{KORRINGA1947392,PhysRev.94.1111} in condensed matter physics, L\"uscher formula  \cite{Luscher:1990ux}  in LCQD  and  Busch-Englert-Rza\.zewski-Wilkens (BERW) formula \cite{Busch98} in a harmonic oscillator trap in nuclear physics community. Other related useful discussions can be found in  e.g. Refs.\cite{Guo_2022_JPG,PhysRevD.103.094520,Guo_2022_JPA,PhysRevC.103.064611}.

Above statement can also be demonstrated  by   the expression of transmission coefficient $T  =  |t |^2 $ for the finite system with $2N+1$ cells,  
\begin{equation}
\frac{1}{T} =1+     {\frac{r_{0}^{r}}{t_{0}}}{\frac{{r_{0}^{l}}^{*}}{{t_{0}}^{*}} \frac{\sin^2(\beta(2N+1)\Lambda)}{\sin^2(\beta \Lambda)}} 
= 1+\bigg(\frac{1}{T_0}-1\bigg)\frac{\sin^2( \beta(2N+1)\Lambda)}{{\sin^2( \beta \Lambda)}} .
\label{TM}
\end{equation}

In addition, Equation (\ref{TM})  shows that there are two distinct cases for which an incident
wave is totally transmitted, i.e. $T=1$. This implies
perfect resonant transmission with no losses and no gain,
regardless of the complex nature of the coupling constants. 

The first case occurs when there is no reflected wave from any individual cell and
this matches the condition when the product of $\frac{r_{0}^{r}}{t_{0}}\frac{{r_{0}^{l}}^{*}}{{t_{0}}^{*}}$ in Eq. (\ref{TM}) is zero (or $T_0=1$). This would lead to the unidirectional
propagation discussed in several studies 
on $\mathcal{P}\mathcal{T}$-symmetric systems,  see, e.g, Refs.~\cite{lin,feng,long1,long2}. This phenomenon is also referred as
the  effect of exceptional points (EPs) that separate the broken and unbroken $\mathcal{P}\mathcal{T}$-symmetric phases, see e.g. Refs.~\cite{10.1038/nphys4323,doi:10.1142/q0178,doi:10.1126/science.aar7709,doi.org/10.1038/s41563-019-0304-9}.

In the second case $\sin (\beta(2N+1)\Lambda)/\sin (\beta \Lambda)=0$. It corresponds to constructive interference
between path reflected from different unit cells at
\begin{equation}
{\beta\Lambda}={\frac{\pi l}{2N+1}}, \qquad |l|=   1,\cdots, N.  \label{res}
\end{equation}
 
In both cases mentioned, we have a perfect
transmission, that is, $ T =1$.
As a consequence, the product of two
the reflection coefficients on the left and right should disappear according to the formula (\ref {ener}). In the case, when one of
reflections reach zero while the other remains non-zero, so-called unidirectional transparency can occur when we have an ideal non-reflective transmission in one direction but not in the other. The experimental demonstration of a unidirectional reflectionless 
at optical 
parity-time metamaterial at optical frequencies is reported in Ref.  \cite{feng}. An outlook on the potential directions and applications 
of controlling sound in non-hermitian acoustic systems can be found in Ref.\cite{gu}.

\subsection{Spectral singularities in periodic system with \texorpdfstring{$2N+1$}{2N+1} cells}

To illustrate the influence of the two factors mentioned above, as well as the role of spectral singularities on the formation of Faraday and Kerr rotations and their shapes let us note that (i)
the spectral singularities arise when both
conditions, $\it Re(\omega) = 0$ and $Im(\omega) = 0$, are satisfied  simultaneously (ii) the location of these poles can be found by solving $1/t(k)=0$.
Based on Eq.(\ref{tandrexpress}),  there are two types of solutions, as was mentioned above:

(i) Type I singularities are given by solutions of  $ \frac{1}{t_0 (\omega)} =0$.  Hence $\cos (\beta \Lambda) =0$ and $\frac{1}{t(\omega)}=0$ are both automatically satisfied:
\begin{equation}
\beta \Lambda=  \pi  l + \frac{\pi}{2}   , \ \ \ \ l\in \mathbb{Z}.
\end{equation}
The type I singularities are originated from a single cell ($N=0$),  and   shared by  the entire system of $2N+1$ cells. The type I solutions  are independent of number of cells and the size of system. The detailed discussion about type I singularities can be found in Sec.\ref{specsingunitcell}.

(ii) type II singularities depend on the size of the system and are given by two conditions,  
\begin{equation}
\cos \left( \beta (2 N+1) \Lambda \right) = 0,   \ \ \ \  Im \left [ \frac{e^{- i k_0 L_0 }}{t_0 (k)} \right ]=0\label{type2}.
\end{equation}
  Hence $\beta \Lambda = \frac{\pi (l+ \frac{1}{2})}{2N+1}$ where $l \in \mathbb{Z}$, above two conditions are  given explicitly by 
  \begin{align}
   \sin \left ( \frac{n \omega_{cr}}{c} L \right )  \left [ \cos \left (\frac{n_0 \omega_{cr}}{c} L_0 \right )    \mathcal{R}(\omega_{cr}) -  \sin \left (\frac{n_0 \omega_{cr}}{c} L_0 \right )     \mathcal{I}(\omega_{cr})  \right]& = \cos \frac{\pi (l + \frac{1}{2})}{2 N+1}  ,  \nonumber \\
 \cos \left (\frac{n_0 \omega_{cr}}{c} L_0 \right )    \mathcal{I}(\omega_{cr}) +  \sin \left (\frac{n_0 \omega_{cr}}{c} L_0 \right )     \mathcal{R}(\omega_{cr})  & = 0.
\end{align} 
At the limit of $V_1\rightarrow 0$, two conditions are reduced to
\begin{equation}
\frac{ V^2_2\omega^2_{\omega_{cr}}}{2c^2nn_0}-\tan \left (\frac{n_0 \omega_{cr}L_0}{c} \right )\cot \left (\frac{n \omega_{cr}L}{c} \right )=\frac{1}{2}\left(\frac{n_0}{n}+\frac{n}{n_0}
\right), \ \ \
\cos\frac{\pi (l+\frac{1}{2})}{2N+1}=
\frac{\cos \left (\frac{n \omega_{cr}L}{c} \right )}{\cos\left (\frac{n_0 \omega_{cr}L_0}{c} \right )}
\label{spectra1}
\end{equation}

\subsection{Large \texorpdfstring{$N$}{N} limit}

As number of cells is increased, all FR and KR angles demonstrate fast oscillating behavior due to $\sin ( \beta (2N+1)\Lambda )$ and $\cos ( \beta (2N+1)\Lambda )$ functions in transmission and reflection amplitudes. These behaviors  are very similar to what happens for tunneling time of a particle through layers of  periodic $\mathcal{P}\mathcal{T}$-symmetric  barriers  that is discussed in Ref.~\cite{peng22}. For the large $N$ systems, we can introduce the  FR and KR angles per unit cell
\begin{align}
\widehat{\theta}^{T/R} (\omega) = \frac{ \theta^{T/R} (\omega)}{(2 N+1) \Lambda} .
\end{align}
The   $N \rightarrow \infty$ limit may be approached by adding a small imaginary part to $\beta$: $\beta \rightarrow \beta + i \epsilon$, where $\epsilon \gg \frac{1}{(2L+1) \Lambda}$.   As discussed in Ref.~\cite{peng22},  adding a small imaginary part to $\beta$ is justified by considering the averaged FR and KR angles per unit cell, which ultimately smooth out the fast oscillating behavior of FR and KR angles.
  Using asymptotic behavior of
\begin{equation}
 \sec(\beta(2N+1)\Lambda)  \propto 2 e^{i \beta(2N+1)\Lambda}  , \ \ \ \   \tan(\beta(2N+1)\Lambda)  \propto 1,
\end{equation}
we find
\begin{equation}
\frac{1}{(2N+1) \Lambda} \ln t (\omega)  \stackrel{ N \rightarrow \infty}{\rightarrow}  i \beta , \ \ \ \  \frac{1}{(2N+1) \Lambda} \ln r (\omega)  \stackrel{ N \rightarrow \infty}{\rightarrow}   0.
\end{equation}
Therefore, as $N \rightarrow \infty$, FR and KR angles per unit cell approach
\begin{equation}
\widehat{\theta}^{T}_1    \stackrel{ N \rightarrow \infty}{\rightarrow} -
  \frac{g}{2n} \frac{\partial  Re[ \beta] }{\partial n} ,  \ \ \ \ 
\widehat{\theta}^{T}_{2}  \stackrel{ N \rightarrow \infty}{\rightarrow} 
 \frac{g}{2 n} \frac{\partial  Im  [\beta]}{\partial n} ,  \ \ \ \  \widehat{\theta}^{R^{r/l}}_{1,2}  \stackrel{ N \rightarrow \infty}{\rightarrow} 0.
\end{equation}
 Also noted that at large $N$ limit,
 \begin{equation}
 \sin(\beta(2N+1)\Lambda)  \propto- \frac{1}{2 i}  e^{-i \beta(2N+1)\Lambda} \stackrel{ N \rightarrow \infty }{ \rightarrow }\infty , 
\end{equation}
   using Eq.(\ref{TM}), one can show that transmission coefficient therefore approaches zero: $T 
 \stackrel{ N \rightarrow \infty}{\rightarrow} 0$.
The relation between $\widehat{\theta}^{T}_2$ and $\widehat{\theta}^{R^{r/l}}_2$ given in Eq.(\ref{theta2RT})  hence is still valid as $N \rightarrow \infty$.

  \begin{figure*}
 \centering
 \begin{subfigure}[b]{0.49\textwidth}
\includegraphics[width=0.99\textwidth]{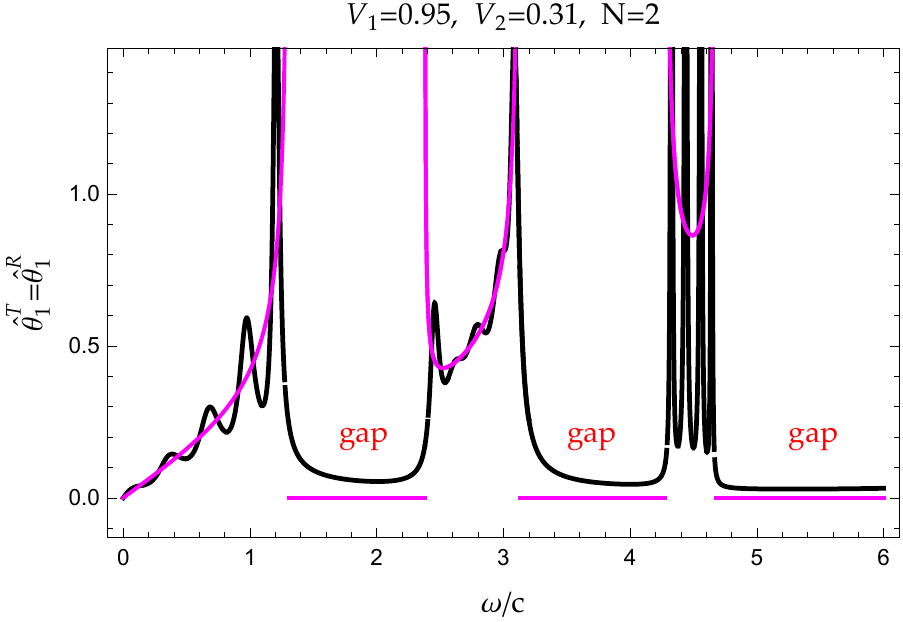}
\caption{   }\label{thetaT1Nplot1}
\end{subfigure} 
\begin{subfigure}[b]{0.49\textwidth}
\includegraphics[width=0.99\textwidth]{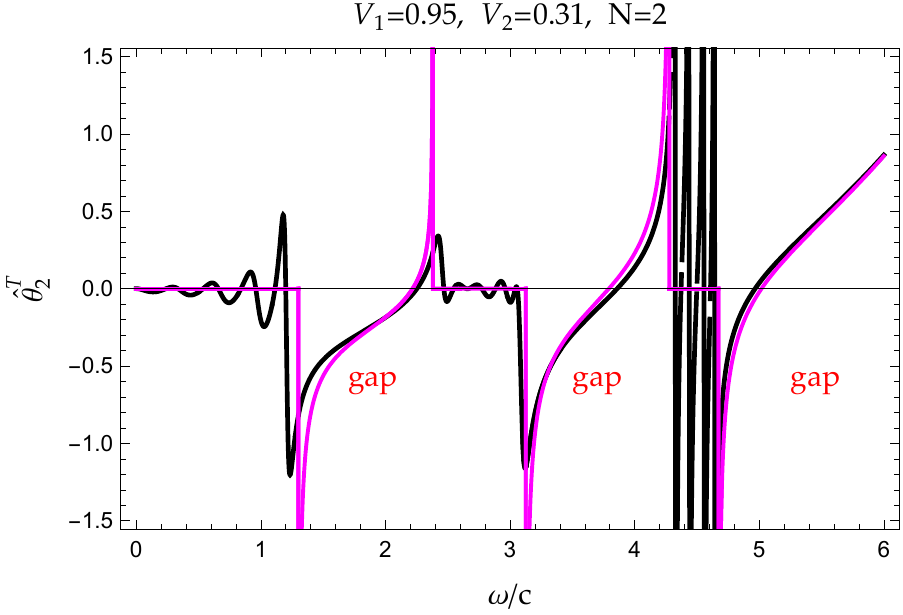}
\caption{    }\label{thetaT2Nplot1}
\end{subfigure}
\begin{subfigure}[b]{0.49\textwidth}
\includegraphics[width=0.99\textwidth]{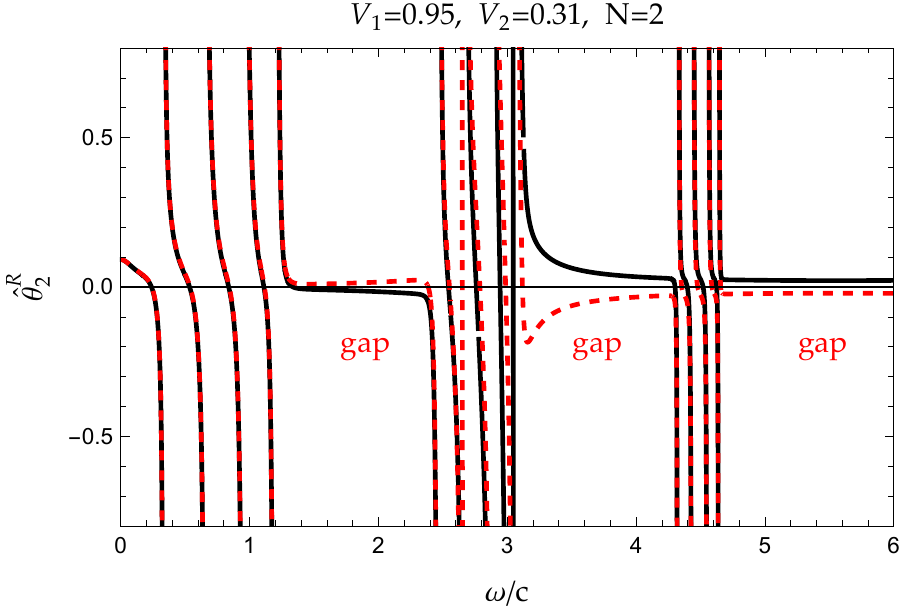}
\caption{ }\label{thetaR2Nplot1}
\end{subfigure}
\begin{subfigure}[b]{0.47\textwidth}
\includegraphics[width=0.99\textwidth]{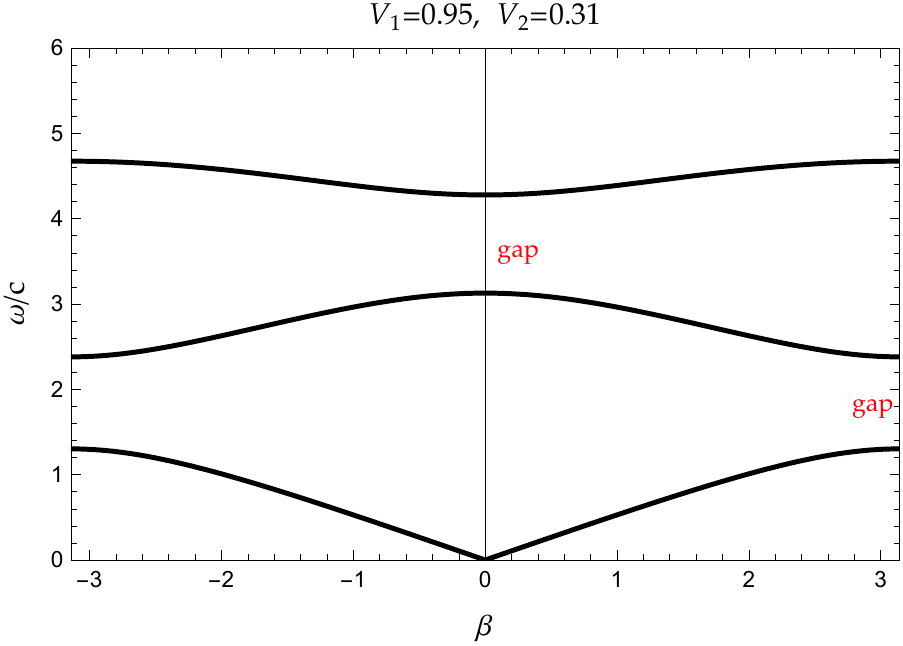}
\caption{ }\label{bandplot1}
\end{subfigure}
\caption{  Plot of   FR and KR angles with $N=2$  vs.  large $N$ limit result: $i \frac{g}{2n} \frac{d \beta}{dn}$   (solid purples):  (a) $\widehat{\theta}^T_1=\widehat{\theta}^R_1$ (solid black) vs. $\frac{g}{2n} \frac{d Re[\beta]}{dn}$  (solid purple);  (b)  $\widehat{\theta}^T_2 $  (solid black) vs. $-\frac{g}{2n} \frac{d Im[\beta]}{dn}$  (solid purple); (c)  $\widehat{\theta}^{R^r}_2 $ (solid black) vs. $\widehat{\theta}^{R^l}_2$ (dashed red); (d)  Band structure plot in unbroken phase. The  parameters are taken as:    $V_1 =0.31 $,  $V_2 =0.95 $,   $L=0.2$, $L_0=0.8$,  $n=2$ and $n_0=1$ }\label{thetaNplots1}  
\end{figure*}

  \begin{figure*}
 \centering
 \begin{subfigure}[b]{0.49\textwidth}
\includegraphics[width=0.99\textwidth]{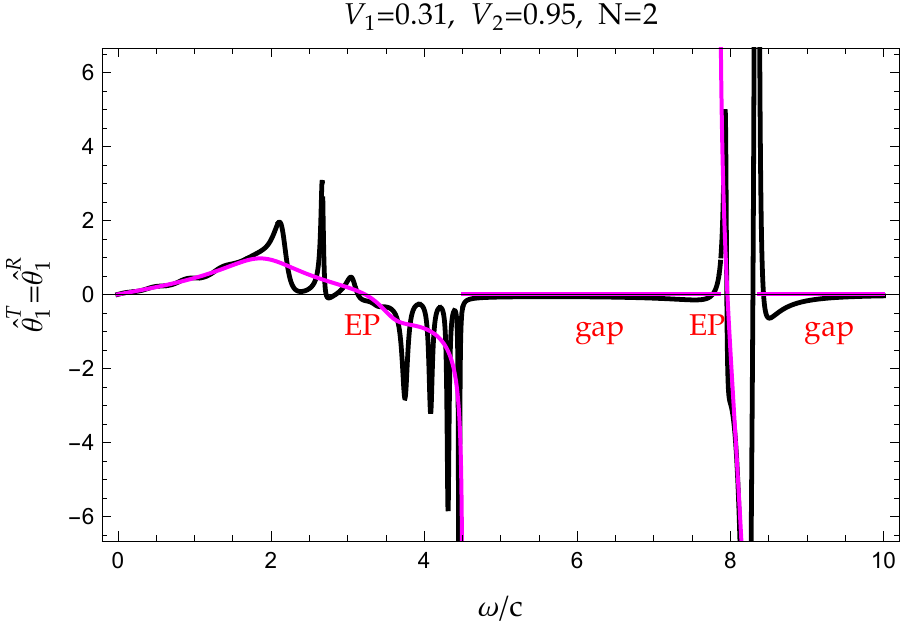}
\caption{   }\label{thetaT1Nplot2}
\end{subfigure} 
\begin{subfigure}[b]{0.49\textwidth}
\includegraphics[width=0.99\textwidth]{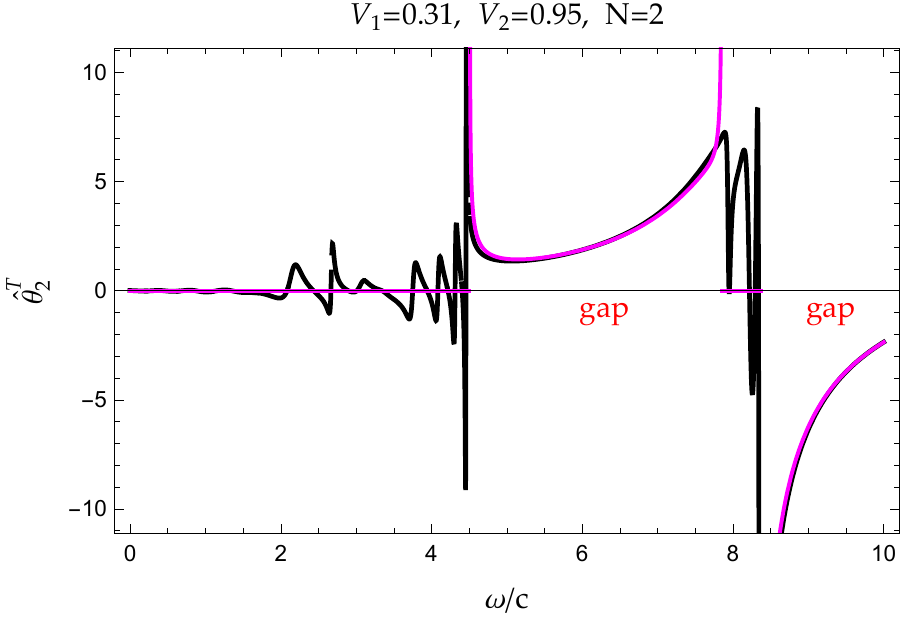}
\caption{    }\label{thetaT2Nplot2}
\end{subfigure}
\begin{subfigure}[b]{0.49\textwidth}
\includegraphics[width=0.99\textwidth]{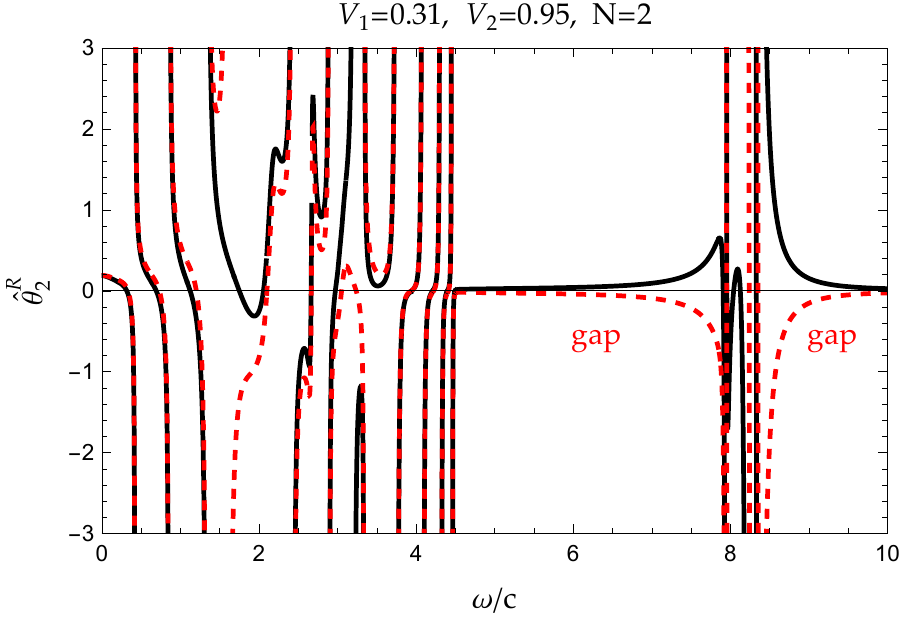}
\caption{  }\label{thetaR2Nplot2}
\end{subfigure}
\begin{subfigure}[b]{0.47\textwidth}
\includegraphics[width=0.99\textwidth]{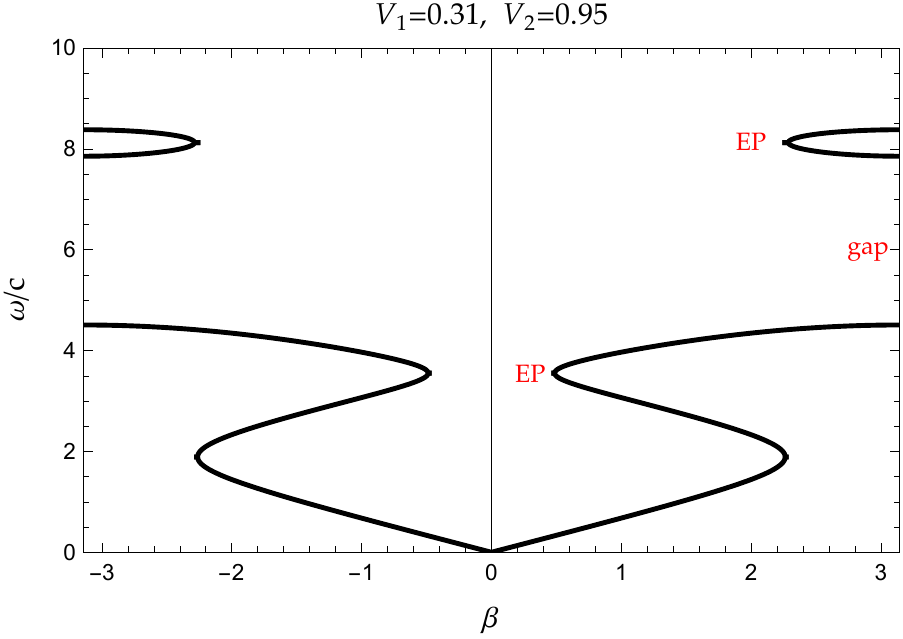}
\caption{ }\label{bandplot2}
\end{subfigure}
\caption{  Plot of   FR and KR angles with $N=2$  vs.  large $N$ limit result: $i \frac{g}{2n} \frac{d \beta}{dn}$   (solid purples): (a) $\widehat{\theta}^T_1=\widehat{\theta}^R_1$ (solid black) vs. $\frac{g}{2n} \frac{d Re[\beta]}{dn}$ (solid purple); (b)  $\widehat{\theta}^T_2 $  (solid black) vs. $- \frac{g}{2n} \frac{d Im[\beta]}{dn}$ (solid purple); (c)  $\widehat{\theta}^{R^r}_2 $ (solid black) vs. $\widehat{\theta}^{R^l}_2$ (dashed red); (d)  Band structure plot in broken phase.  The  parameters are taken as:    $V_1 =0.31 $,  $V_2 =0.95 $,   $L=0.2$, $L_0=0.8$, $n=2$ and $n_0=1$. }\label{thetaNplots2}  
\end{figure*}

The examples of FR and KR angles per unit cell for  a$\mathcal{P}\mathcal{T}$-symmetric finite system with five cells  are shown in Fig.~\ref{thetaNplots1} and Fig.~\ref{thetaNplots2}, compared with the large $N$ limit results. As we can see in Fig.~\ref{thetaNplots1} and Fig.~\ref{thetaNplots2}, the  $\theta^{T}_1$ and $\theta^{T}_2$ angles oscillating around the  large $N$ limit results. Even for the small size system, we can see clearly that the   band structure of infinite periodic system is already showing up.  The oscillating KR angles are consistent with zero at large $N$ limit. In addition, in broken $\mathcal{P}\mathcal{T}$-symmetric phase in Fig.~\ref{thetaNplots2}, EPs can be visualized even for a small size system, where two neighbouring bands merge and the  $\mathcal{P}\mathcal{T}$ becomes totally transparent:  both $\theta^{T}_1$ and $\theta^{T}_2$ approach zero.

For a real refractive index profile, the sign of $\theta^{T}_1$ is always positive due to the fact that $\theta^{T}_1$ is closely related to the density of states. However, in $\mathcal{P}\mathcal{T}$-symmetric systems, $\theta^{T}_1$ is now associated with a generalized density of states, which can be either positive or negative, see discussion in Ref.\cite{peng22,PhysRevResearch.4.023083}. In this sense
the negative spike(s) in Fig.~\ref{thetaNplots1} and Fig.~\ref{thetaNplots2} around the some frequencies provide the formal justification of the existence of such negative states. Turning negative of $\theta^{T}_1$ is closely related to the motion of poles across the real axis moving from unphysical sheet (the second Riemann sheet) into physical sheet (the first Riemann sheet), for more details see Refs.\cite{peng22,GGJ-2022}. 
Since $\theta^{T}_1$ ($\theta^{R}_1$) is assumed to be related to the density of states, it is natural that it is practically zero in all forbidden bands and takes a giant leap to a very large number at the end of each band.

\section{Discussion and summary}\label{summary}

In summary, we studied the anomalous behavior of the Faraday (transmission) and polar Kerr (reflection) rotation angles of the propagating light, in finite periodic parity-time ($\mathcal{P}\mathcal{T}$) symmetric structures, containing $2N+1$ cells.

We have obtained closed
form expressions for FR and KR angles for a single cell consisting of two complex $\delta$-potentials placed on both boundaries of the ordinary dielectric slab.
It is shown that, for a given set of parameters describing the system, a phase transition-like anomalous behavior of Faraday and Kerr rotation angles in a parity-time symmetric systems can take place.
In the anomalous phase the value of one of the Faraday and Kerr rotation angles can become negative, and both angles suffer from spectral singularities and give a strong enhancement near the singularities. It is shown that due to symmetry constraints, the real part of the complex angle of KR, $\theta^{R}_1$, is always equal to the $\theta^{T}_1$ of FR, no matter what phase the system is in. The imaginary part of KR angles  $\theta^{R^{r/l}}_2$ are also related to the $\theta^{T}_2$ of FR by parity-time symmetry.

We find that, in the limit of weak scattering, the Kerr and Faraday rotation angles increase linearly with the length of the system. In this approximation the effects of multiple
reflections within the layers are not significant.
We have also shown, based on the modified Kramers-Kronig relations, that only the three angles FR and KR are completely independent.

\acknowledgments

P.G. and V.G.   acknowledge support from the Department of Physics and Engineering, California State University, Bakersfield, CA.  V.G., A.P-G. and E.J. would like to thank UPCT for partial financial support through the concession of "Maria Zambrano ayudas para la recualificación del sistema universitario español 2021-2023" financed by Spanish Ministry of Universities with financial funds "Next Generation" of the EU.

\appendix

\section{Determinant Approach}\label{detapproach}

This section is devoted to more mathematical interest. We combine two non-perturbative  approaches, that sufficiently
completely describe of photon (electron)  behaviour in a random potential to study the energy spectrum and scattering matrix elements in the  $\mathcal{P}\mathcal{T}$ system without actually determining the photon eigenfunctions.

In both approaches, the Green's function was calculated exactly for two different models. 
In the first model, we are dealing with
the sum of $\delta$-potentials distributed randomly with an arbitrary strength. The second model was used to calculate the passage of a free particle through a layered
system, which is characterized by random parameters of the layers.

A convenient formalism to study one dimensional scattering
systems satisfying the stationary Schr\"odinger equation or the Helmholtz equation relevant to optical Bragg grating is developed in Ref \cite{aronov91,VG90}. The approach allows one to express the transmission
and reflection amplitudes of a wave propagating in a one-dimensional random layered  
structure through the
characteristic determinant $D_N$ ($N$ is the number of the boundaries), which depends on the amplitudes of reflection of a single scatter
only. The transmission amplitude $t_N$ of waves through the systems 
can be presented in the form    
\begin{equation}
t_N=\frac{e^{ik|x_N-x_1|}}{D^{0}_N}, \label{t1a}
\end{equation}
where the characteristic determinant $D^{0}_N$ reduces to a recursive equation that is convenient for both numerical and analytical approaches.

This paper presents a generalization of the determinant approach to the case of $\mathcal{P}\mathcal{T}$-symmetric (non-symmetric) systems consisting of $(N - 1)$ dielectric multilayers with two delta potentials in each. The detailed and,
in many respects, complete description and analysis of the Faraday and Kerr  effects in such a system discussed. Specifically, our investigations focus on the 
periodic finite size diatomic $\mathcal{P}\mathcal{T}$-symmetric model.
We predict that for a given set of parameters describing the system the Faraday and Kerr rotation angles show a non-trivial transition with a change in sign. 
In the anomalous phase the value of one of the Faraday and Kerr rotation angles can become negative, and both angles suffer from spectral singularities and give a strong enhancement near the singularities. 

Let us consider $(N - 1)$ dielectric multilayer system labeled $n ={1, \cdots, N - 1}$ between two semi-infinite
media. The positions of the boundaries of the $nth$ dielectric layer, characterized by the constant $\epsilon_n$, are given by $x_n$ and $x_{n+1}$
respectively. 
The left and right ends of the system are at $x=x_{_N}$ and $x=x_{1}$ with $\epsilon_0 =\epsilon_{N}$, respectively.
We assume that a plain EMW wave is incident from the left (with the dielectric permittivity $\epsilon_0$) onto the boundary at $x = x_1$
and evaluate the amplitude of the reflected wave and the wave propagating in the semi-infinite media for $x > x_N$, characterized by $\epsilon_N$. In the further discussion we will assume, that the first and last layers of
the multilayer system make interfaces with the vacuum.

We also assume that we know the transmission $t_{n,n+1}$ and reflection amplitudes (from the left $r_{n,n+1}$ and the right $r_{n+1,n}$) of the EMW from a  single $Z_n\delta(x-x_n)$ scatter, located at the contact of two semi-infinite media I and II at $x=x_n$. Using the results of the transmission and reflection amplitudes 
for the single scatter, we will build 
characteristic determinant $D_N$ for $N$ scatters and obtain the total transmission $t_N$ and reflection amplitudes $r^{N}_{L}$ and $r^{N}_R$.  
The transmission amplitudes from left and from right equal each other are given by
\begin{equation}
t_{n,n+1}= t_{n+1,n}\equiv \frac {2\sqrt{\frac{k_n}{k_{n-1}}}}{1+\frac{k_n}{k_{n-1}}-i\frac{\gamma}{k_{n-1}}Z_n},  \ \ \ \ k_n=\frac{\omega}{c} n , \ \ \ \ \gamma \equiv \left(
\frac{\omega}{c} \right)^2. \label{t21B}
\end{equation}
Similarly,
\begin{equation}
r_{n,n+1}=\frac {1-\frac{k_{n+1}}{k_{n}}+i\frac{\gamma}{k_{n}}Z_n}{1+\frac{k_{n+1}}{k_{n}}-i\frac{\gamma}{k_{n}}Z_n}, \ \ \ \ 
r_{n+1,n}=\frac{\frac{k_{n+1}}{k_{n}}-1+i\frac{\gamma}{k_{n}}Z_n}{1+\frac{k_{n+1}}{k_{n}}-i\frac{\gamma}{k_{n}}Z_n}. \label{r21Ar21B}
\end{equation}
We can easily verify by using Eq.(\ref{t21B}) and Eq.(\ref{r21Ar21B}) that 
the conservation law is satisfied, provided that $Z_n$ is real:
\begin{equation}
t_{n,n+1}t^{*}_{n,n+1}+r_{n,n+1}r^{*}_{n,n+1}=\frac{4\frac{k_n}{k_{n-1}}+(1-\frac{k_n}{k_{n-1}})^2+(\frac{\gamma}{k_{n-1}}Z_n)^2}{(1+\frac{k_n}{k_{n-1}})^2+(\frac{\gamma}{k_{n-1}}Z_n)^2}=1
\end{equation}
In the case of a complex value $Z_n$, the conservation law cannot hold, since the system is not $\mathcal{P}\mathcal{T}$-symmetric and can be described
by only complex energy eigenvalue. Later, when we "build" the characteristic determinant $D_N$ for the entire system with $N$ complex potentials, distributed arbitrary, we will return to the conservation law of the system in more detail.

 Assuming that we know the explicit expression for the amplitude of reflection from a single-scattering delta potential, see Eq.(\ref{r21Ar21B}), we now turn to a closer investigation of the system with two complex potentials.
 Following Refs.\cite{aronov91}, we can present the determinant $D_2$ of two delta potentials located at points $x_1$ and $x_2$ 
($L$ = $x_2$- $x_1$) on the left and right boundaries of a dielectric slab surrounded by two semi-infinite media with permittivities $\epsilon_0$ (left) and $\epsilon_2$ (right), respectively. The dielectric slab itself is characterized by permittivity $\epsilon_ 1$. The explicit form of $D_2$ is
\begin{equation}
D^{0}_2=\frac{1}{(1+r_{21})(1+r_{32})}\det D_2,
\label{det2}
\end{equation}
where 
\begin{equation}
\begin{aligned} 
\det D_2\equiv\begin{vmatrix}
1 & r_{23}e^{ik_1(x_2-x_1)} \\
r_{21}e^{ik_1(x_2-x_1)} & 1
\end{vmatrix},
\label{det2a}
\end{aligned}
\end{equation}
and $r_{n,n+1}$ is given by Eq.(\ref{r21Ar21B}) 
with the appropriate choice of $n$ and $Z_n$.
Let us add another boundary from the right, at the point $x_3$, i.e. we consider a layered heterostructure
consisting of two films with permittivities $\epsilon_1$ and $\epsilon_2$,
placed between two semi-infinite media $\epsilon_0$ and $\epsilon_3$. 

Next, adding another delta complex potential $Z_3$ at $x_3$ 
the new $D_3$, which now is $3 \times 3$ determinant, 
can be written
as
\begin{equation}
\begin{aligned}
D^{0}_3=&\prod^{3}_{l=1}\frac{1}{(1+r_{l+1,l})}\det D_3
\label{det3a}
\end{aligned}
\end{equation}
where
\begin{equation}
\begin{aligned}
\det D_3 \equiv \begin{vmatrix}
1 & r_{23}e^{ik_1(x_2-x_1)} & r_{34}e^{ik_1(x_2-x_1)}e^{ik_2(x_3-x_2)}\\
r_{21}e^{ik_1(x_2-x_1)} & 1& r_{34}e^{ik_2(x_3-x_2)}\\
r_{21}e^{ik_1(x_2-x_1)}e^{ik_2(x_3-x_2)}&r_{32}e^{ik_2(x_3-x_2)}&1\\
\end{vmatrix}.
\label{det3aa}
\end{aligned}
\end{equation}
By continuing adding new boundaries and complex potential $Z_n$
at the points $x_4$,\ldots,
$x_N$, we will obtain an $N$-multilayer system, each layer of which contains two delta potentials. 
This system will be characterized by the product of $N$ by $N$ determinant $D_N$
\begin{equation}
D^{0}_N=\prod^{N}_{l=1}\frac{1}{(1+r_{l+1,l})}\det  D^{N}_{l,n},\label {DN}
\end{equation}
with the following matrix elements $D^{N}_{l,n}$:
\begin{equation}
D^{N}_{l,n}=\left\{
\begin{array}{l l}
\delta_{ln}+(1-\delta_{ln})r_{l,l-1}e^{ik_l|x_l-x_n|}, \ \  l\ge n, \\
    \delta_{ln}+(1-\delta_{ln})r_{l-1,l}e^{ik_l|x_l-x_n|}, \ \  n\ge l.\\
 \end{array} \right. \label{nm}
\end{equation}
The characteristic determinant $D_N$ can be presented as a
determinant of a Teoplitz tridiagonal matrix that satisfies the following recurrence relationship:
$$
D_N = A_ND_{N-1}-B_ND_{N-2},
$$
where $D_{N-1}$ ($D_{N-2}$) is the determinant equation (\ref{nm}) with the
$Nth$ and also the $(N-1)th$
 row and column omitted.
 The initial conditions for the recurrence relations are
$D_0 = 1$, $D_{-1}=0$, 
$D_1\equiv A_{1}=1$.  The coefficients $A_N$, $B_N$ can be obtained from the explicit
form of $D^{N}_{n,l}$ (see Eq. (\ref{nm})).
For $N>
1$ we have
$$A_n =1+ \frac{r_{n,n+1}}{r_{n-1,n}}(1+r_{n-1,n}+
r_{n,n-1})e^{2ik_{n}|x_n-x_{n-1}|}=1+B_n-
r_{n,n-1}r_{n,n+1}e^{2ik_{n}|x_n-x_{n-1}|}
$$
and
$$B_n = 
\frac{r_{n,n+1}}{r_{n-1,n}}(1+r_{n,n-1})(1+
r_{n-1,n})e^{2ik_{n}|x_n-x_{n-1}|}
$$

In concluding, let us stress once more that Eqs.(\ref{DN}) and (\ref{nm}) 
may be viewed as generalization of the characteristic determinant method that can be applied to the Helmholtz  (Shr\"odinger) equation with complex potentials, distributed arbitrary and find scattering matrix elements without actually determining the photon (electron) eigenfunctions.

\bibliography{biblio.bib}

\end{document}